\documentclass[9pt,twocolumn,twoside]{osajnl}

\usepackage{xcolor}

\journal{ol} % Choose journal (ao, aop, josaa, josab, ol, pr)

\setboolean{shortarticle}{true}
\title{Partial quantum revivals of localized condensates in distorted  lattices}

\author[1,2]{Dogyun Ko}
\author[1,2]{Meng Sun}
\author[1,2]{Alexei Andreanov}
\author[3]{Y. G. Rubo}
\author[1,2]{I. G. Savenko}

\affil[1]{Center for Theoretical Physics of Complex Systems, Institute for Basic Science (IBS), Daejeon 34126, Korea}
\affil[2]{Basic Science Program, Korea University of Science and Technology (UST), Daejeon 34113, Korea}
\affil[3]{Instituto de Energ\'{\i}as Renovables, Universidad Nacional Aut\'onoma de M\'exico, Temixco, Morelos 62580, Mexico}

\begin{abstract}
We report on a peculiar propagation of bosons loaded by a short Laguerre-Gaussian pulse in a nearly flat band of a lattice potential. 
Taking a system of exciton-polaritons in a kagome lattice as an example, we show that an initially localized condensate propagates in a specific direction in space if anisotropy is taken into account. 
This propagation consists of quantum jumps, collapses, and revivals of the whole compact states, and it persists given any direction of anisotropy. This property reveals its signatures in the tight-binding model and, surprisingly, it is much more pronounced in a continuous model. 
Quantum revivals are robust to the repulsive interaction and finite lifetime of the particles. 
Since no magnetic field or spin-orbit interaction is required, this system provides a new kind of easily implementable optical logic. 
\end{abstract}

\setboolean{displaycopyright}{true}

\begin{document}

\maketitle

\section{Introduction}
Specially prepared, coherent states~\cite{perelomovbook86} of simple quantum systems, as a harmonic oscillator or a spin, exhibit nondestructive dynamics along classical trajectories, but, in general, the wave packets spread out in time. 
This spreading reverses if the energies of the system are rational numbers, leading to full or fractional revivals of the quantum states as, e.g., in the case of the Brown states~\cite{brown73} of the hydrogen atom~\cite{PhysRevA.42.6308}. 

Quantum transport properties change drastically for particles loaded in a flat band (FB) of a periodic potential~\cite{Derzhko:2015aa}. 
In particular, in an ideal case described by the tight-binding model, the particle can occupy only a few neighboring sites of the lattice, forming the compact localized state (CLS)~\cite{Sutherland:1986aa}. 
The CLS is an eigenstate of the single-particle Hamiltonian, and thus it does not propagate nor does it spread in time. 
In down-to-earth situation, however, there exist many factors hindering this perfect localization, including only approximate flatness of the band, finite lifetime of particles, and interparticle interactions. 
These relentless obstacles lead us to the natural questions: Are FBs actually feasible? How do reasonably small deformations of a FB affect the dynamics of initially perfect CLSs? And can one at all speak about any (quasi-)CLSs in realistic circumstances?

We address these issues by studying the dynamics of exciton polaritons (later polaritons) loaded in a nearly perfect FB. 
Polaritons are typical composite bosons arising due to strong coupling between the microcavity photons and quantum well excitons~\cite{kavokin17} 
%and represent new hybrid eigen modes of the system. 
and experiencing dissipation and possessing finite lifetime in semiconductor microcavities. However, under external coherent or incoherent excitation they might condense~\cite{kasprzak06, balili07}, manifesting macroscopic occupation of a particular state---phenomenon also referred to as polariton lasing. 
The state in question can be localized or trapped, and there exist multiple experimental scenarios of polariton trapping, either in a single state or in a periodic network. In particular, polaritons can be exposed to spatially-periodic acoustic waves~\cite{cerdamendez10} or created in etched microcavities. 
%cerdamendez13
Polariton condensation in artificial periodic potentials has recently become a remarkably active field of research. 
Polaritons develop new transport properties if loaded in honeycomb~\cite{jacqmin14}, kagome~\cite{masumoto12,gulevich16}, or 1D and 2D Lieb~\cite{baboux16, klembt17, whittaker18} lattices, occasionally forming topologically protected~\cite{karzig15,nalitov15,bardyn15,stjean17,chunyanli18} and single-particle quasi-flat bands (which we will also refere to as FBs in what follows). 
%\alexei{AA: Most examples of topological flatbands are only quasi FB, and I feel we need to specify that.}
%

Several questions remain open and disputable in this field. In particular, polariton condensates in FBs possess a rather short coherence length, and it is unclear if this is the consequence of a disorder, or condensate fragmentation is a generic property of out-of-equilibrium systems loaded in a FB.
However, a periodic long-range order can appear spontaneously in resonantly-driven cavities~\cite{gavrilov18}.
Moreover, condensed out-of-equilibrium particles should not necessarily occupy the lowest energy state (typically the $\Gamma$ point). There emerge $\pi$-condensates (at the edge of a band) in 1D potentials~\cite{lai07} and $d$-condensates in 2D square lattices~\cite{kim11}. The choice of the phase of the condensate is controlled by the polariton-polariton interaction, which can also lead to space-time intermittency regime in microcavities with periodic potentials (lattices)~\cite{yoon19}.

Recently, the kagome lattice has been a subject of intense research thanks to its remarkable properties. In particular, it represents one of the paradigmatic systems, where frustration and destructive interference take place. In addition, due to the presence of FB and CLSs, interacting bosons at low density are predicted to form a supersolid phase, which manifests itself in periodic density alterations~\cite{huber10}, and the many-body ground state of high-density bosons in the kagome FB is currently a hot topic of research~\cite{maiti19}.  
The nearly-localized states %(which we will also refer to as CLSs in what follows) 
in kagome lattice flat band have been observed experimentally~\cite{Zong:16}. 
Compactness of these states makes them promising candidates for manipulation of optical quantum information. 
Our work addresses the experimentally important question on what happens to these states in the case of a nonperfect flat band. We show that there is no simple spreading and decay of CLSs, and deformation of the lattice can lead to nontrivial quantum revivals, making this system promising platform for optical logic manipulations.

\section{Results and discussion}
\begin{figure}[b!]
\includegraphics[width=0.47\textwidth]{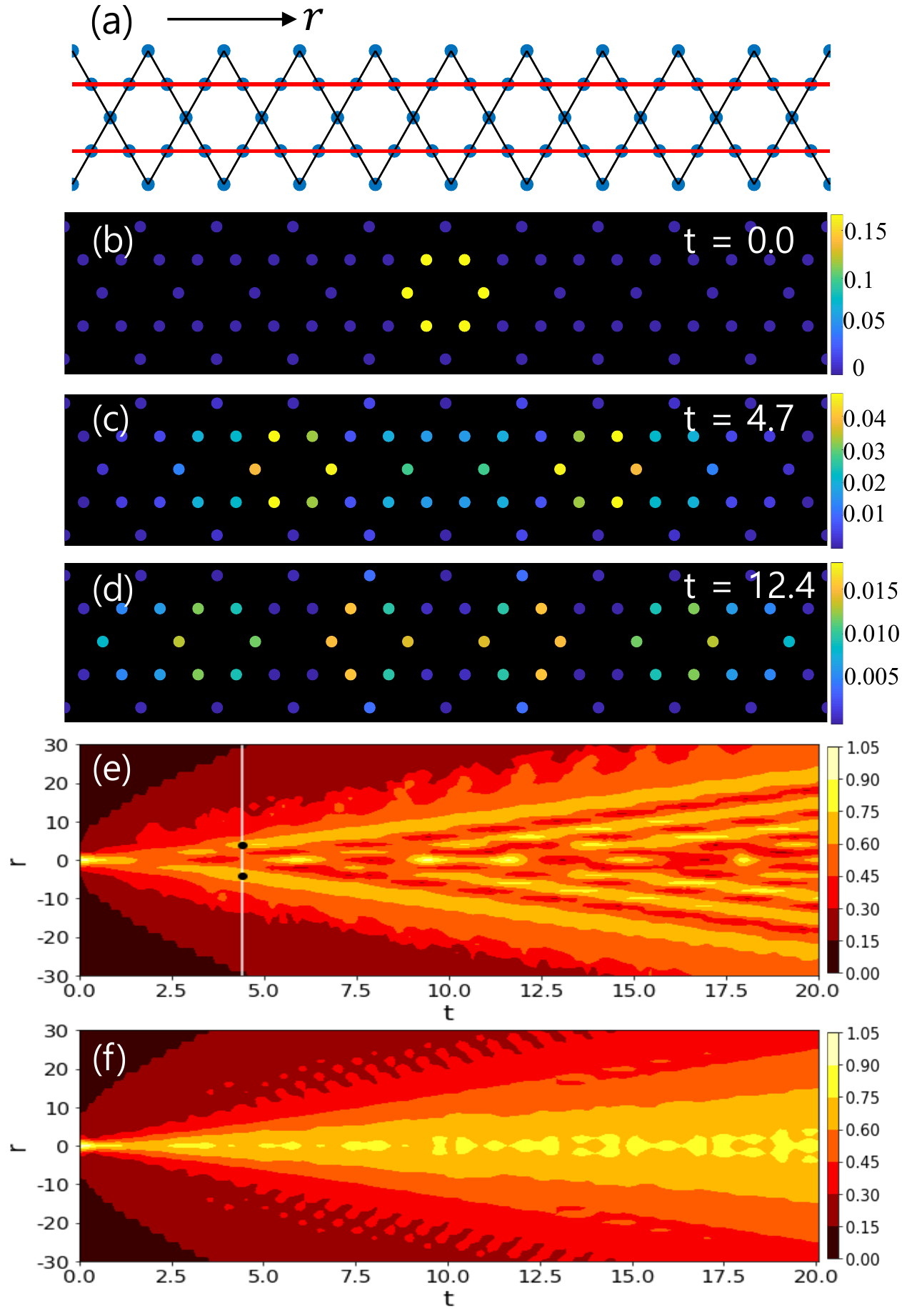}
\caption{Tight-binding description. (a) \textcolor{black}{A kagome chain, with red (horizontal) bonds having half the hopping integrals of the other (black) bonds.}
State of the system: (b) $t=0$ (arb. units), initial state is the exact CLS; (c) $t=4.7$; (d). $t=12.4$. (e-f)  Density plot of CLS revivals evolution in time with a kagome chain and kagome lattice, respectively: every point of the y-axis is a hexagon. The color  denotes the value of the overlap with a CLS. The vertical line in (e) indicates the time shown in (c) and the black dots represent the two copies of the original CLS. }
\label{fig1}
\end{figure}

%\textbf{\emph{Tight-binding description.}} 
We start by using a regular tight-binding model on a kagome chain. 
It has a FB in the case of uniform hoppings. 
We lift the FB degeneracy by altering the hopping coefficients in horizontal bonds [see Fig.~\ref{fig1}(a)] and study the unitary evolution of the system initialized in one of the CLSs, which occupies sites of a single hexagon [Fig.~\ref{fig1}(b)].
\textcolor{black}{Namely, the hoppings on the horizontal bonds, indicated by the red color in Fig.~\ref{fig1}(a), have half the value of the non-horizontal bonds. This mimics the lattice distortion of the continuous model discussed below. Changing the ratio of the hopping coefficients mainly affects the period of the revivals and the spreading velocity.}
%We modify the hopping parameters for dilated neighbouring bonds, shown in  Fig.~\ref{fig1}(a).} 
%\textcolor{green}{YR: All red bond are dilated? I do not understand Fig.1(a).}
 
%

Due to the distortion, the CLS ceases to be an exact eigenstate and it spreads over the chain with time. 
We observe a signature of revivals of the CLS at short times [Fig.~\ref{fig1}(c)], and nonperfect twins of the original CLS that appear at shifted positions. 
\textcolor{black}{We note that this is a partial revival, since the whole state of the quantum system does not fully coincide with its original state, and it is characterized by the presence of a few CLS condensates further away from the creation center.}
Eventually the CLS is completely destroyed [Fig.~\ref{fig1}(d)]. 
This destruction occurs faster for stronger distortions.

Figures~\ref{fig1}(e) and~\ref{fig1}(f) illustrate the full process of the CLS partial revivals and their eventual destruction in (e) a kagome chain, when we leave a single layer of hexagons composed of triangles, and (f) full 2D kagome lattice.
We plot the overlap of every hexagon in the chain with a CLS and time evolution of the overlaps.
One observes rather clear signatures of revivals at short times, but they are gradually suppressed at longer times.
Similar but weaker signatures  also occur in a 2D kagome lattice, where a stripe of sites with different couplings is introduced. 
The revivals of the initial CLS in the tight-binding model are weaker than the ones seen in the continuous model, that we discuss next.

% \begin{figure}[h]
%     \centering
%     \begin{subfigure}[b]{0.4\textwidth}
%         \includegraphics[width=\textwidth]{lattice_phase.pdf}
%         \label{fig:2}
%     \end{subfigure}
%     \captionsetup{justification=centerlast,singlelinecheck=false}
%     \caption{Schematics of the system. (a). the 2D Kagome lattice. A compact localized states(CLS) can be created at the center of the lattice in six pillars. Blues are quantum wells and each pillars are touched each other. (b). band structure of the Kagome lattice shows 2 lowest dispersive bands and the highest nondispersive band. (c). To create CLS at the center, Laguerre Gaussian pumping is used. (d). Phase of LG beam}
% \end{figure}
%
%
%
\begin{figure}[b!]
\centering
\includegraphics[width=0.45\textwidth]{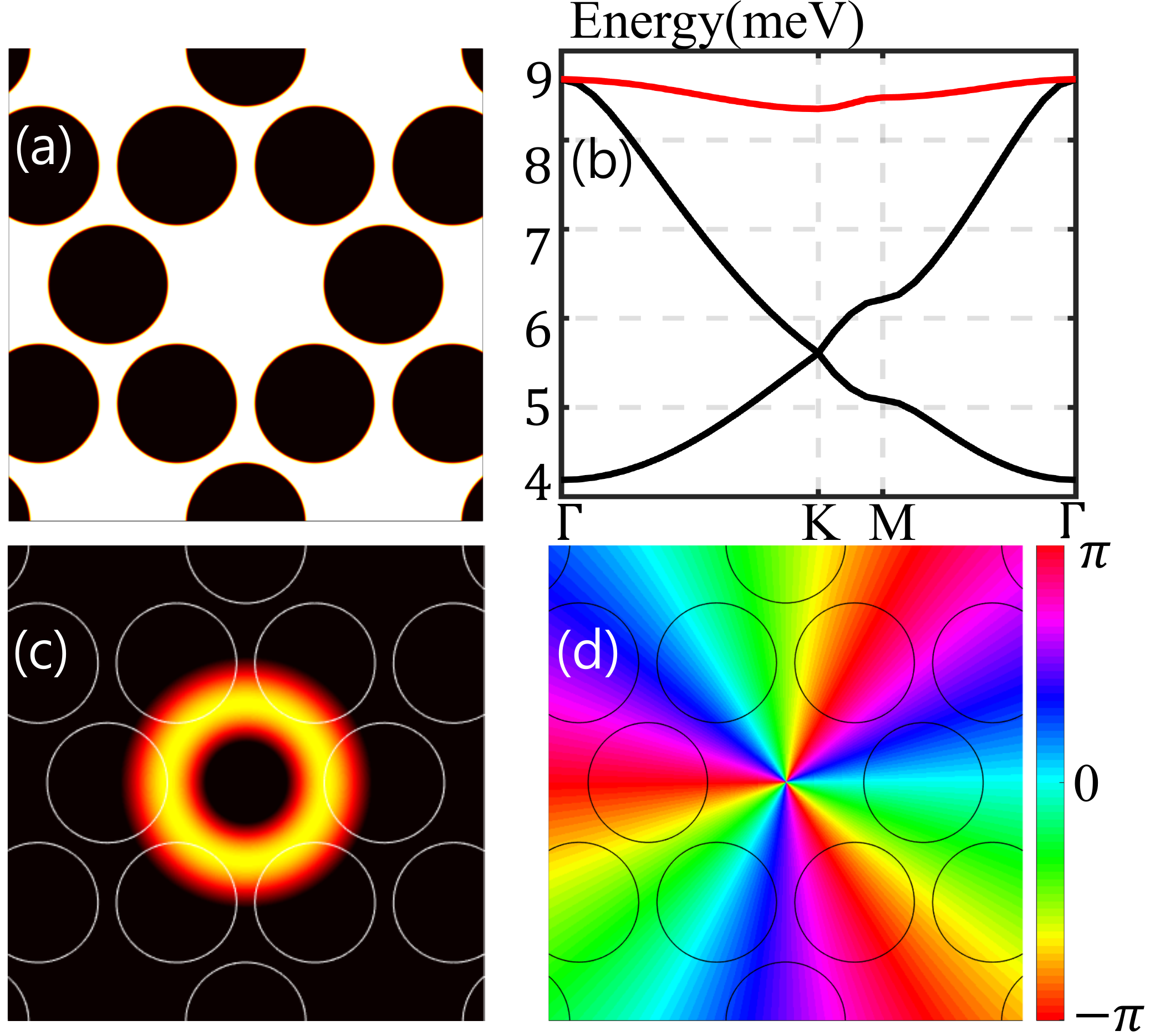}
\caption{System schematic. (a) A 2D Kagome lattice. A compact localized state can be created at the center of the lattice, occupying six pillars. 
(b) The band structure, showing two lowest dispersive bands and the highest non-dispersive one. 
(c) The intensity of Laguerre-Gaussian (LG) pumping, that aligns well with the hexagonal structure of the kagome lattice. 
(d). Phase of the LG beam.}
\label{fig2}
\end{figure}
%
%
%

%--------------------
%--------------------
%--------------------

%\textbf{\emph{Continuous model.}} 
We study the formation and propagation of polaritons coupled with an incoherent excitonic reservoir, using the evolution equations
\begin{eqnarray}
\mathrm{i}\hbar\frac{\partial\psi}{\partial t}&=&\mathcal{H}\psi+\mathcal{P}-\frac{\mathrm{i}\hbar}{2\tau_{p}}\psi+\frac{\mathrm{i}\hbar G n_{R}}{2}\psi+\alpha|\psi|^{2}\psi+\textcolor{black}{gn_{R}\psi}
\label{eq:pl}\\
\frac{\partial n_{R}}{\partial t}&=&P_{in}-\frac{n_{R}}{\tau_{R}}-G|\psi|^{2}n_{R}, 
\label{eq:rev}
\end{eqnarray}
where $\psi$ is a polariton macroscopic wave function, $\mathcal{H} = -\frac{\hbar^2}{2m_p}\nabla^2 +V$ is the Hamiltonian, responsible for the propagation of the particles with the effective mass $m_p$ in a kagome lattice potential $V$, shown in Fig.~\ref{fig2}(a);
$\mathcal{P}$ is the coherent pumping term; $\tau_p$ and $\alpha$ are the polariton lifetime and polariton-polariton interaction constant, respectively;
\textcolor{black}{g is polariton-reservoir interaction constant;}
$n_R$ is the reservoir particle density, which is incoherently pumped by the $P_{in}$ term and has a finite lifetime $\tau_R$; the reservoir and polaritons are coupled by a phenomenological constant $G$. 

Taking $V=0$ inside the lattice sites (pillars) and $V=30$ meV outside the pillars, we first solve the eigenvalue problem $\mathcal{H}\psi=E\psi$ in the framework of the continuous model, and we find the band structure, shown in Fig.~\ref{fig2}(b).
We note that, unlike the tight-binding model, the third band here is only approximately flat.
To create a CLS of the \textcolor{black}{nearly} flat band, we employ a coherent Laguerre-Gaussian (LG) pump at the center of the lattice, as shown in Fig.~\ref{fig2}(c), and which is given by
\begin{equation}
    \mathcal{P}=P_{0}\left(\frac{r}{R}\right)^l L^{0}_{l}\left(\frac{r^2}{R^2}\right) exp\left[-\left(\frac{r^2}{R^2}\right)+\mathrm{i}(l\phi-\omega_{0}t)\right],
    \label{eq:LG}
\end{equation}
where $P_{0}$ is the pump amplitude, $r$ is the radial distance from the center of the plaquette, $\phi$ is the azimuth angle, $R=0.7\,\mu\mathrm{m}$ is the radius of the pumping ring, $l=3$ determines the phase difference between neighboring lattice sites (adjacent pillars), which in our case is $\pi$ [see Fig.~\ref{fig2}(d)]. 
Furthermore, $\omega_{0}$ is the pump frequency, which we put equal to the frequency of the third band of the kagome lattice at the $\Gamma$ point.
\textcolor{black}{For short pulses, the spreading in energy is greater than the width of the flat band, so that the exact coincidence of frequencies is not necessary for the CLS excitation.}

\textcolor{black}{We note, that it is convenient to excite CLSs in kagome lattice with the Laguerre-Gaussian pump, as compared to the Lieb lattice~\cite{PhysRevB.98.161204}, due to the absence of parasite pumping of the sites outside the CLS, i.e., the sites, where the population should be zero as a result of destructive interference.}

We apply this coherent pump for a short time, thus making a short pulse, of about $0.5\,\mathrm{ps}$ duration, to generate the CLS and let it evolve with the support of the continuous background incoherent pumping $P_{in}$.
\textcolor{black}{
In the absence of $P_{in}$ the population of the system quickly decreases. The subthreshold pumping $P_{in}<P_{th}=(G\tau_{R}\tau_{P})^{-1}$ helps to prolong the existence of the condensate, by compensating the losses due to the finite lifetime.
Nevertheless, this background incoherent pump does not hinder the revival phenomenon, as we show below.
}

\textcolor{black}{It is important, that the main excitation comes from the resonant Laguerre-Gaussian pulse. We assume this pulse alone to be strong enough to create substantial macroscopic occupation of CLS, such that its evolution can be described by mean-field equation even without additional non-resonant pumping. The non-resonant pumping improves applicability of the Gross-Pitaevskii equation by additional feeding of this macroscopic state. It is assumed that the number of particles is still macroscopic after revivals, so that the semiclassical description remains valid. 
On the other hand, it is necessary to keep the background non-resonant pumping below the threshold. Otherwise, the effect would be masked by the spontaneous condensate formation.}

\begin{figure}[t!]
\centering
\includegraphics[width=0.45\textwidth]{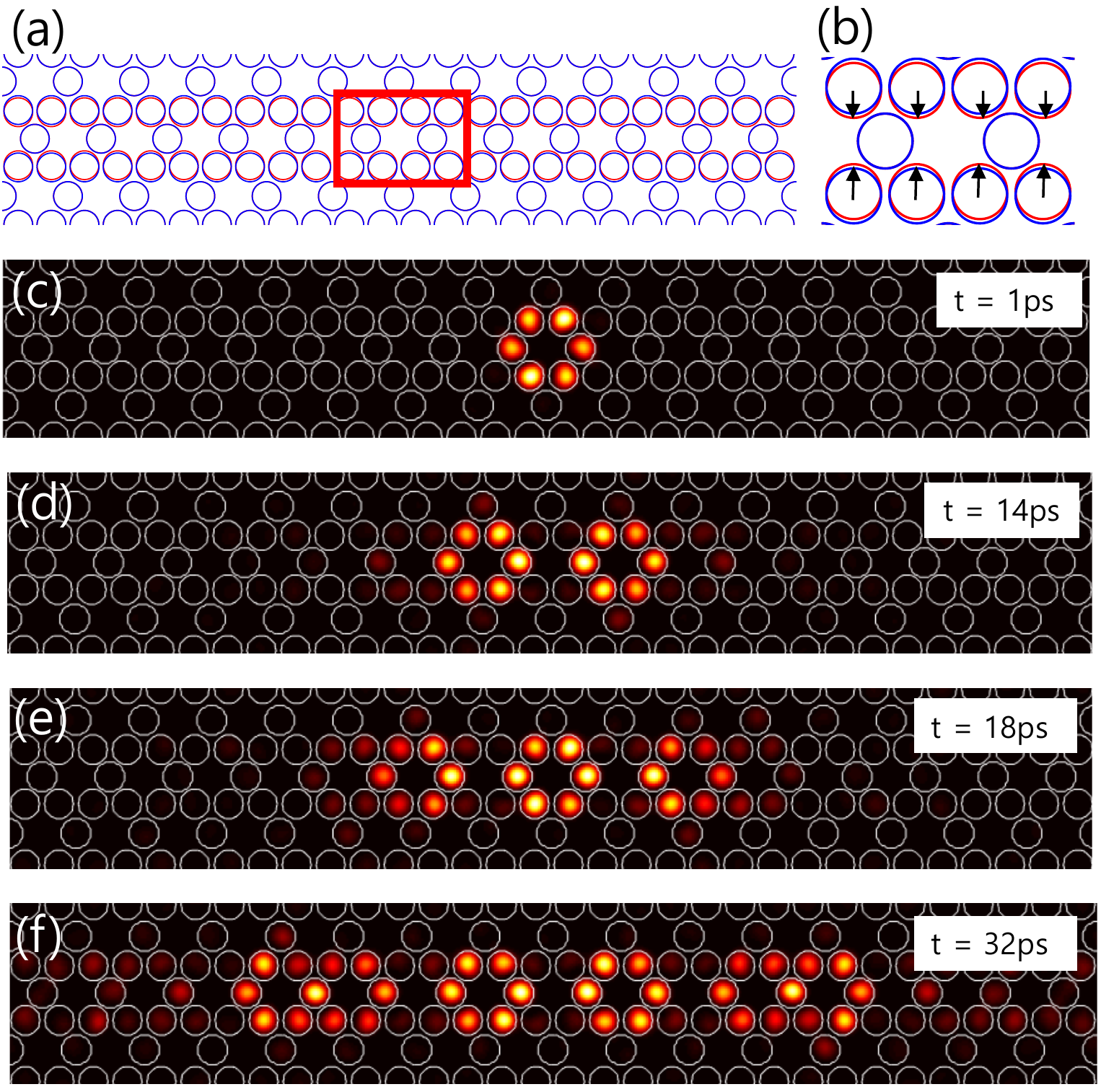}
\includegraphics[width=0.45\textwidth]{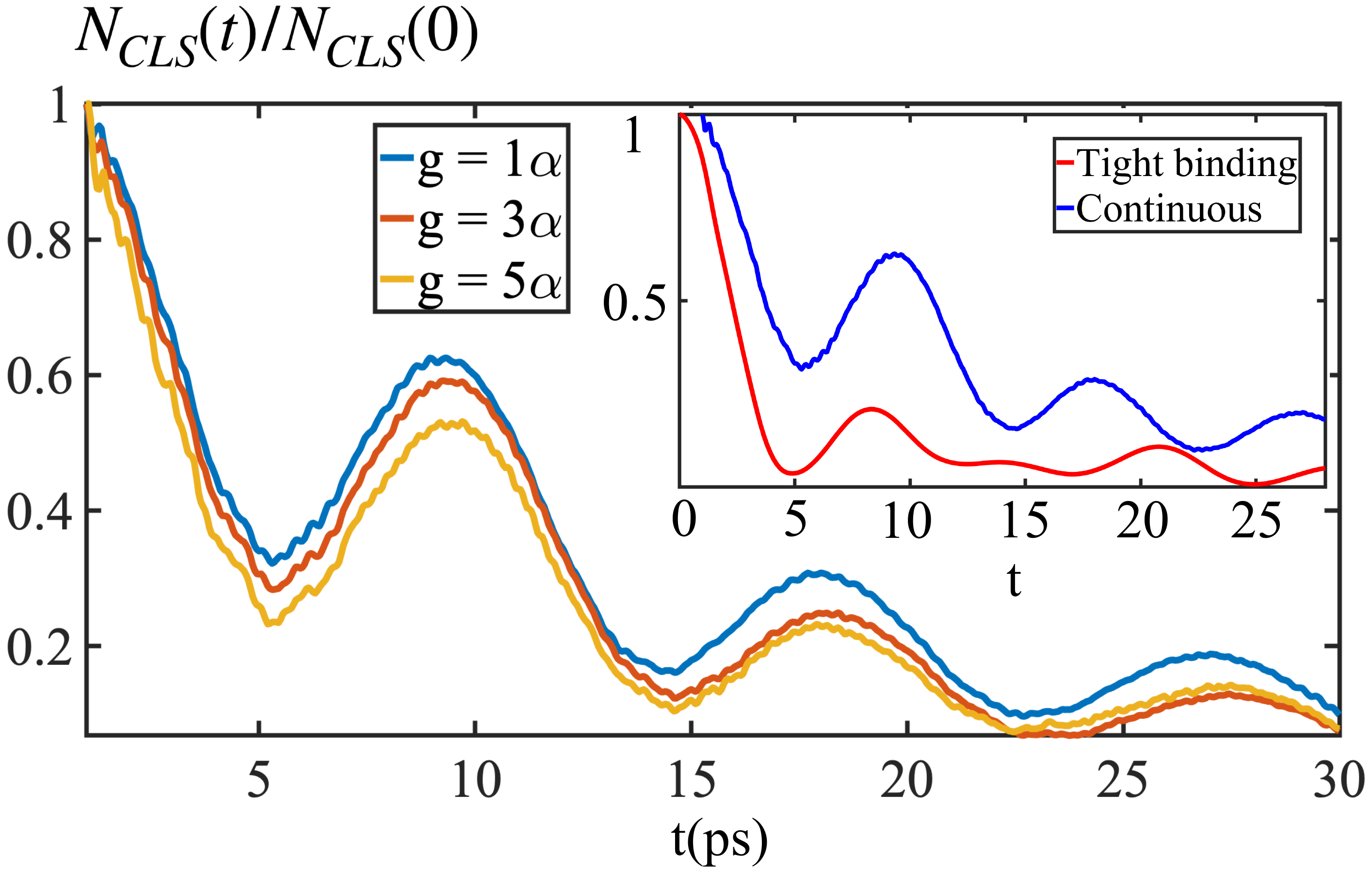}
\caption{(a) The kagome $y$-squeezed strip. (b) Red pillars are squeezed vertically. Snapshots of CLS at 1~ps (c), 14~ps (d), 18~ps (e), and 32~ps (f). \textcolor{black}{(see Visualization 1)}. 
The vertical lines indicate times, shown in (d-f). (g) Temporal evolution of the number of particles in CLS at the center of the strip \textcolor{black}{for different polariton-reservoir interaction constants}: manifestation of the partial revivals. 
\textcolor{black}{(Inset) Comparison of the time evolution of the CLS weight in the tight-binding (red) and continuous (blue) model.} 
}
    \label{fig4}
\end{figure}

%\textbf{\emph{Partial jumps and revivals.}} 
Similarly to the tight-binding case, we introduce anisotropy in the system, as it is shown in Fig.~\ref{fig4}, and solve Eqs.~(\ref{eq:pl}) and (\ref{eq:rev}) numerically. 
As a result, we observe (i) quasi-1D propagation of the CLSs in certain directions, which can be seen as partial jumps, and (ii) revivals of the CLSs at their original position and at different positions in space. 

We study the kagome strips squeezed in $y$-direction, as shown in Figs.~\ref{fig4}(a,b) 
%\textcolor{black}{(the results of the $x$-squeezing are discussed in Appendix)}.
\textcolor{black}{We note that experimentally there exist two methods to squeeze the lattice. One way is to etch the lattice with a defective line, where the distances are slightly different. Another way is to apply unidirectional mechanical stress to the lattice, which will cause effective deformation of the lattice. 
In this work we simulate the former method.}
%, and Fig.~\ref{fig4}(b)}blue pillars stand for the non-deformed kagome lattice, while the red pillars are shifted vertically.
\textcolor{black}{The lattice constant is 3$\mu\mathrm{m}$. The radius of the pillars is $0.65\mu\mathrm{m}$ and the pillars are shifted as $0.1\mu\mathrm{m}$.}
The revival of the CLS on a kagome $y$-squeezed strip is shown in Fig.~\ref{fig4}(c-f). The LG beam shines in the center of the lattice for $0.5$ ps. Figure~\ref{fig4}(c) shows the particle density when the coherent pump is switched off. 
The condensate now propagates both to the left and to the right, and the CLS at the center of the strip completely disappears at about $14\,\mathrm{ps}$. 
At the same time there appear new CLSs next to the center, see Fig.~\ref{fig4}(d).
Further on, the condensate continues to propagate and creates new CLSs in neighboring sites, but the CLS at the center is partially restored at $32\,\mathrm{ps}$, see Fig.~\ref{fig4}(e).
Figure~\ref{fig4}(f) shows that the CLS at the center of the strip disappears again and the condensate reaches the boundary of the system.
\textcolor{black}{Figure~\ref{fig4}(g) shows the influence of the interaction constant $g$ on revivals. 
By increasing $g$, the number of particles in the strip decreases.
This figure also proves that the revival phenomenon still occurs with higher polariton-reservoir interaction.}
% the dynamics of the ratio of the CLS polaritons at the central plaquette of the lattice and the total number of polaritons for different polariton-reservoir

\textcolor{black}{In this work, we considered a clean system and disregarded any structural disorder. Typical potential disorder in polariton lattices of 0.2 meV is much smaller than the height 30 meV of the potential barriers in the kagome lattice. Thus we do not expect strong suppression of the revivals.
We note also, that the effect has an essentially local character: the particles created by the initial pulse spread and come back involving a few lattice sites only. Therefore, experimentally one can try to find a place in the lattice with particularly weak disorder to observe the revivals.
}
%{On the other hand, it would be interesting to understand the interplay of the disorder and the FB properties in this setup.}

The most prominent feature of revivals is that new CLSs inherit and bequeath the shape of the original CLS. The memory is also conserved in the phases of the wave functions of CLSs, which separate after another collapse and reunite at later time. 
This phenomenon does not require the transport through the edge states, or external magnetic field, or some particular type of spin-orbit coupling, and still the particles show spectacular quantum propagation, keeping the entrusted memory.

\textcolor{black}{We expect the effect to persist in other lattices with a destroyed FB. 
For other lattices, the revivals might be absent if, e.g., the FB is protected by a symmetry and the deformations respect the symmetry, as in the case of the deformed Lieb lattice.}
%
%
%
%
%
%Furthermore, the design which we propose in this work can be easily implemented in an experiment since pillared polariton microcavities can be routinely produced. 
% YR: I replaced "unidirectional" by "spectacular quantum". Unidirectional which means either to the left or to the right, as the particle propagates along the edge states or in the Hall effect. In our case it seems to be misleading: the mirror symmetry is unbroken and the CLS propagates both ways.

%--------------------
%--------------------
%--------------------

\section{Conclusions}
We reported on a peculiar unexpected dynamics of bosonic condensates loaded in a distorted kagome lattice, where the transport of particles manifests itself in jumps and partial quantum revivals of compact localized states. 
In the tight-binding model, the dynamics of the initial CLS takes place by its reappearance in the neighboring sites, but the revival of the original CLSs is not observed. In contrast, in the continuous model, the deformation of the nearly flat band by squeezing the lattice in either horizontal or vertical direction leads to visible revivals of the polariton wave packet. 
%The kagome $y$-squeezed strips show better revivals than the kagome $x$-squeezed strips, in addition, polaritons in $x$-squeezed strips propagate faster. 
\textcolor{black}{The physical mechanism of the revival phenomenon is the constructive interference of waves reflected back from the lattice while propagating uni-dimensionally.} 
This effect can be additionally supported by a background pump, opening new possibilities for developing all-optical logical elements based on polariton condensates in nearly flat bands.
\textcolor{black}{We expect that our work will stimulate the research towards building new quantum devices, such as kagome optomechanical
lattice~\cite{Wan:17}. Moreover, exciton-polariton delocalization and revivals can be used for information transfer and storage~\cite{cite-key}.}
%We believe that is due to the additional pillars between the plackets in Kagome Y strips.
%The effect of condensate revival can be prolonged by considering the additional incoherent sub-threshold feed of the polariton system. 
%We should find another method to create the revival without anisotropy. We only have shown fractional revival on the Kagome strips. However, we also speculate that another types of Kagome strips can show perfect revival to the original state. We hope this results can encourage other researches to investigate other types of revival on the different lattices.

\section{Acknowledgements}
D.K. and M.S. contributed equally to this work. 
We thank S.~Flach and C.~Danieli for useful discussions 
and acknowledge the support of the Institute for Basic Science in Korea (Project No.~IBS-R024-D1). YGR acknowledges the support from CONACYT (Mexico) under the Grant No.\ 251808 \textcolor{black}{and by PAPIIT-UNAM Grant No.\ IN106320}.

\section{Disclosures}
The authors declare no conflicts of interest.

\vspace{-\baselineskip}
\bibliography{main}

\begin{thebibliography}{10}
\newcommand{\enquote}[1]{``#1''}

\bibitem{perelomovbook86}
A.~Perelomov, \emph{Generalized Coherent States and Their Applications}
  (Springer-Verlag, Berlin Heidelberg, 1986).

\bibitem{brown73}
L.~S. Brown, {\protect\JournalTitle{American Journal of Physics}} \textbf{41},
  525 (1973).

\bibitem{PhysRevA.42.6308}
Z.~D. Gaeta and C.~R. Stroud, {\protect\JournalTitle{Phys. Rev. A}}
  \textbf{42}, 6308 (1990).

\bibitem{Derzhko:2015aa}
O.~Derzhko, J.~Richter, and M.~Maksymenko, {\protect\JournalTitle{International
  Journal of Modern Physics B}} \textbf{29}, 1530007 (2015).

\bibitem{Sutherland:1986aa}
B.~Sutherland, {\protect\JournalTitle{Phys. Rev. B}} \textbf{34}, 5208 (1986).

\bibitem{kavokin17}
A.~V. Kavokin, J.~J. Baumberg, G.~Malpuech, and F.~P. Laussy,
  \emph{Microcavities} (Oxford University Press, 2017).

\bibitem{kasprzak06}
J.~Kasprzak, M.~Richard, S.~Kundermann, A.~Baas, P.~Jeambrun, J.~M.~J. Keeling,
  F.~M. Marchetti, M.~H. Szyma{\'n}ska, R.~Andr{\'e}, J.~L. Staehli, V.~Savona,
  P.~B. Littlewood, B.~Deveaud, and L.~S. Dang, {\protect\JournalTitle{Nature
  (London)}} \textbf{443}, 409 (2006).

\bibitem{balili07}
R.~Balili, V.~Hartwell, D.~Snoke, L.~Pfeiffer, and K.~West,
  {\protect\JournalTitle{Science}} \textbf{316}, 1007 (2007).

\bibitem{cerdamendez10}
E.~A. Cerda-M\'endez, D.~N. Krizhanovskii, M.~Wouters, R.~Bradley, K.~Biermann,
  K.~Guda, R.~Hey, P.~V. Santos, D.~Sarkar, and M.~S. Skolnick,
  {\protect\JournalTitle{Phys. Rev. Lett.}} \textbf{105}, 116402 (2010).

\bibitem{jacqmin14}
T.~Jacqmin, I.~Carusotto, I.~Sagnes, M.~Abbarchi, D.~D. Solnyshkov,
  G.~Malpuech, E.~Galopin, A.~Lema\^{\i}tre, J.~Bloch, and A.~Amo,
  {\protect\JournalTitle{Phys. Rev. Lett.}} \textbf{112}, 116402 (2014).

\bibitem{masumoto12}
N.~Masumoto, N.~Y. Kim, T.~Byrnes, K.~Kusudo, A.~L{\"o}ffler, S.~H{\"o}fling,
  A.~Forchel, and Y.~Yamamoto, {\protect\JournalTitle{New J. Phys.}}
  \textbf{14}, 065002 (2012).

\bibitem{gulevich16}
D.~R. Gulevich, D.~Yudin, I.~V. Iorsh, and I.~A. Shelykh,
  {\protect\JournalTitle{Phys. Rev. B}} \textbf{94}, 115437 (2016).

\bibitem{baboux16}
F.~Baboux, L.~Ge, T.~Jacqmin, M.~Biondi, E.~Galopin, A.~Lemaitre,
  L.~Le~Gratiet, I.~Sagnes, S.~Schmidt, H.~E. Tureci, A.~Amo, and J.~Bloch,
  {\protect\JournalTitle{Phys. Rev. Lett.}} \textbf{116}, 066402 (2016).

\bibitem{klembt17}
S.~Klembt, T.~H. Harder, O.~A. Egorov, K.~Winkler, H.~Suchomel, J.~Beierlein,
  M.~Emmerling, C.~Schneider, and S.~H{\"o}fling, {\protect\JournalTitle{Appl.
  Phys. Lett.}} \textbf{111}, 231102 (2017).

\bibitem{whittaker18}
C.~E. Whittaker, E.~Cancellieri, P.~M. Walker, D.~R. Gulevich, H.~Schomerus,
  D.~Vaitiekus, B.~Royall, D.~M. Whittaker, E.~Clarke, I.~V. Iorsh, I.~A.
  Shelykh, M.~S. Skolnick, and D.~N. Krizhanovskii,
  {\protect\JournalTitle{Phys. Rev. Lett.}} \textbf{120}, 097401 (2018).

\bibitem{karzig15}
T.~Karzig, C.-E. Bardyn, N.~H. Lindner, and G.~Refael,
  {\protect\JournalTitle{Phys. Rev. X}} \textbf{5}, 031001 (2015).

\bibitem{nalitov15}
A.~V. Nalitov, D.~D. Solnyshkov, and G.~Malpuech, {\protect\JournalTitle{Phys.
  Rev. Lett.}} \textbf{114}, 116401 (2015).

\bibitem{bardyn15}
C.-E. Bardyn, T.~Karzig, G.~Refael, and T.~C.~H. Liew,
  {\protect\JournalTitle{Phys. Rev. B}} \textbf{91}, 161413(R) (2015).

\bibitem{stjean17}
P.~St-Jean, V.~Goblot, E.~Galopin, A.~Lema{\^\i}tre, T.~Ozawa, L.~Le~Gratiet,
  I.~Sagnes, J.~Bloch, and A.~Amo, {\protect\JournalTitle{Nat. Photon.}}
  \textbf{11}, 651 (2017).

\bibitem{chunyanli18}
C.~Li, F.~Ye, X.~Chen, Y.~V. Kartashov, A.~Ferrando, L.~Torner, and D.~V.
  Skryabin, {\protect\JournalTitle{Phys. Rev. B}} \textbf{97}, 081103(R)
  (2018).

\bibitem{gavrilov18}
S.~S. Gavrilov, {\protect\JournalTitle{Phys. Rev. Lett.}} \textbf{120}, 033901
  (2018).

\bibitem{lai07}
C.~W. Lai, N.~Y. Kim, S.~Utsunomiya, G.~Roumpos, H.~Deng, M.~D. Fraser,
  T.~Byrnes, P.~Recher, N.~Kumada, T.~Fujisawa, and Y.~Yamamoto,
  {\protect\JournalTitle{Nature (London)}} \textbf{450}, 529 (2007).

\bibitem{kim11}
N.~Y. Kim, K.~Kusudo, C.~Wu, N.~Masumoto, A.~L{\"o}ffler, S.~H{\"o}fling,
  N.~Kumada, L.~Worschech, A.~Forchel, and Y.~Yamamoto,
  {\protect\JournalTitle{Nat. Phys.}} \textbf{7}, 681 (2011).

\bibitem{yoon19}
S.~Yoon, M.~Sun, Y.~G. Rubo, and I.~G. Savenko, {\protect\JournalTitle{Phys.
  Rev. A}} p. in press (2019).

\bibitem{huber10}
S.~D. Huber and E.~Altman, {\protect\JournalTitle{Phys. Rev. B}} \textbf{82},
  184502 (2010).

\bibitem{maiti19}
S.~Maiti and T.~Sedrakyan, {\protect\JournalTitle{Phys. Rev. B}} \textbf{99},
  174418 (2019).

\bibitem{Zong:16}
Y.~Zong, S.~Xia, L.~Tang, D.~Song, Y.~Hu, Y.~Pei, J.~Su, Y.~Li, and Z.~Chen,
  {\protect\JournalTitle{Opt. Express}} \textbf{24}, 8877 (2016).

\bibitem{PhysRevB.98.161204}
M.~Sun, I.~G. Savenko, S.~Flach, and Y.~G. Rubo, {\protect\JournalTitle{Phys.
  Rev. B}} \textbf{98}, 161204(R) (2018).

\bibitem{Wan:17}
L.-L. Wan, X.-Y. L\"{u}, J.-H. Gao, and Y.~Wu, {\protect\JournalTitle{Opt.
  Express}} \textbf{25}, 17364 (2017).

\bibitem{cite-key}
L.-L. Wan, X.-Y. L{\"u}, J.-H. Gao, and Y.~Wu,
  {\protect\JournalTitle{Scientific Reports}} \textbf{7}, 15188 (2017).

\end{thebibliography}
\bibliographyfullrefs{main}

%
%
%
%

%\pagebreak

%\section{Appendix: The kagome x-squeezed strip}

\end{document}